\documentclass[12pt,english]{article}
\usepackage[T1]{fontenc}
\usepackage[latin9]{inputenc}
\usepackage{geometry}
\usepackage{babel}
\usepackage{amsmath}
\usepackage{amssymb}
\usepackage{esint}
\usepackage[unicode=true,pdfusetitle,
 bookmarks=true,bookmarksnumbered=false,bookmarksopen=false,
 breaklinks=false,pdfborder={0 0 1},backref=false,colorlinks=false]
 {hyperref}

\makeatletter
\newcommand{\lyxaddress}[1]{
\par {\raggedright #1
\vspace{1.4em}
\noindent\par}
}

\usepackage{graphicx}
\usepackage{xcolor}
\definecolor{darkblue}{rgb}{0,0,0.5} 
\usepackage{transparent}
\renewcommand\[{\begin{equation}}
\renewcommand\]{\end{equation}}

\makeatother

\begin{document}

\title{Continuum Approach to Non-equilibrium Quantum Functional Integral}

\author{Filippo Bovo}

\maketitle

\lyxaddress{School of Physics and Astronomy, University of Birmingham, Edgbaston,
Birmingham, B15 2TT, UK.}
\begin{abstract}
Standard derivations of the functional integral in non-equilibrium
quantum field theory are based on the discrete time representation
\cite{Kamenev2011,AltlandSimons2010Book}. In this work we derive
the non-equilibrium functional integral for non-interacting bosons
and fermions using a continuum time approach by accounting for the
statistical distribution through the boundary conditions and using
them to evaluate the Green's function.
\end{abstract}
Non-equilibrium quantum field theory has its foundations on the closed
time contour. The contour, introduced by Schwinger \cite{Schwinger1960,Schwinger1961},
is used to evaluate expectation values and replaces the real time
axis of the equilibrium theory. This contour was developed in statistical
quantum mechanics by Kadanoff and Baym \cite{KadanoffBaym1962} and
its formulation was made simpler and more transparent by Keldysh \cite{Keldysh}.
Keldysh's formulation is even more transparent in its functional integral
analogue, developed by Kamenev \cite{Kamenev2005}. This formulation,
used also in Refs. \cite{Kamenev2011,AltlandSimons2010Book}, accounts
for the statistical distribution as an off-diagonal component of the
inverse Green's function in the discrete time representation. The
Green's function is calculated by matrix inversion and the continuum
limit is eventually taken. Instead, in this work we adopt a continuum
approach: we derive the functional integral by accounting for the
statistical distribution through the boundary conditions and we evaluate
the Green's function as a solution of a differential equation with
boundary conditions derived from those of the functional integral.

The central object of non-equilibrium quantum functional integral
is the partition function along the closed time contour,

\begin{equation}
Z=\frac{\mathrm{Tr}\left[\hat{\mathcal{U}}\left(t_{i},t_{f}\right)\hat{\mathcal{U}}\left(t_{f},t_{i}\right)\hat{\rho}\right]}{\mathrm{Tr}\left[\hat{\rho}\right]}\,,\label{eq:Partition_Function_0}
\end{equation}
where $\hat{\rho}$ is the initial distribution, $\hat{\mathcal{U}}(t_{f},t_{i})=\mathbb{T}e^{-\textrm{i}\int_{t_{i}}^{t_{f}}\mathrm{d}t\hat{H}(t)}$
($\hbar=1$) is the time-evolution operators from the initial time
$t_{i}$ to the final time $t_{f}$ and $\mathbb{T}$ is the time-order
operator \cite{Kamenev2011}. The operators $\hat{\mathcal{U}}\left(t_{f},t_{i}\right)$
and $\hat{\mathcal{U}}\left(t_{i},t_{f}\right)$ are referred to as
forward and backward time-evolution operators \cite{Kamenev2011}.
This partition function has meaning only if there are different external
sources on forward and backward time-evolution operators. These sources
are used to generate correlation functions and are usually set to
zero at the end of a calculation \cite{Kamenev2011}. When the external
sources are the same on forward and backward time-evolution operators,
$\hat{\mathcal{U}}\left(t_{i},t_{f}\right)\hat{\mathcal{U}}\left(t_{f},t_{i}\right)=1$
and the partition function is identically $Z=1$. However, since the
following derivation is independent of sources, we omit them, keeping
in mind that they should be present.

We start by constructing the functional integral representation of
$Z$ for a non-interacting single-level system of energy $\varepsilon$
populated by bosons or fermions and we will extend it to many-level
systems at the end. The Hamiltonian is $\hat{H}(t)=\varepsilon\hat{a}^{\dagger}\hat{a}$,
where $\hat{a}$ and $\hat{a}^{\dagger}$ are bosonic or fermionic
creation and annihilation operators, satisfying the canonical commutation
relation $[\hat{a},\hat{a}^{\dagger}]_{+}=1$ for bosons or anti-commutation
relation $[\hat{a},\hat{a}^{\dagger}]_{-}=1$ for fermions. To derive
the functional integral representation of the partition function,
we consider the coherent states of bosons and fermions, $\phi$ and
$\bar{\phi}$, respectively associated to $\hat{a}$ and $\hat{a}^{\dagger}$
through the relations $\hat{a}\bigl|\phi\bigr\rangle=\phi\bigl|\phi\bigr\rangle$
and $\bigl\langle\phi\bigr|\hat{a}^{\dagger}=\bar{\phi}\bigl\langle\phi\bigr|$,
where $\phi$ and $\bar{\phi}$ are complex numbers for bosons or
Grassmann number for fermions. Inserting the resolution of identity
of coherent states between the three operators in the numerator of
Eq. (\ref{eq:Partition_Function_0}) and using the coherent state
representation of the trace, the partition function becomes,
\begin{equation}
\begin{aligned}Z=\frac{1}{\mathrm{Tr}\left[\hat{\rho}\right]} & \int\mathrm{d}\bigl[\bar{\phi}_{i}^{-},\phi_{i}^{-}\bigr]\mathrm{d}\bigl[\bar{\phi}_{i}^{+},\phi_{i}^{+}\bigr]\mathrm{d}\bigl[\bar{\phi}_{f},\phi_{f}\bigr]\,e^{-\bar{\phi}_{i}^{+}\phi_{i}^{+}}\bigl\langle\phi_{i}^{+}\bigr|\hat{\rho}\bigl|\zeta\phi_{i}^{-}\bigr\rangle\times\\
 & e^{-\bar{\phi}_{i}^{-}\phi_{i}^{-}}\bigl\langle\phi_{i}^{-}\bigr|\hat{\mathcal{U}}\left(t_{i},t_{f}\right)\bigl|\phi_{f}\bigr\rangle\,e^{-\bar{\phi}_{f}\phi_{f}}\bigl\langle\phi_{f}\bigr|\hat{\mathcal{U}}\left(t_{f},t_{i}\right)\bigl|\phi_{i}^{+}\bigr\rangle\,,
\end{aligned}
\label{eq:Partition_Function_1}
\end{equation}
where $\mathrm{d}\bigl[\bar{\phi},\phi\bigr]=\mathrm{d}(\mathrm{Re}\phi)\,\mathrm{d}(\mathrm{Im}\phi)/\pi$,
$\zeta=+1$ for bosons and $\mathrm{d}\bigl[\bar{\phi},\phi\bigr]=\mathrm{d}\bar{\phi}\,\mathrm{d}\phi$,
$\zeta=-1$ for fermions. We used the superscript $+$ and $-$ for
initial states acting respectively on forward and backward time-evolution
operators. As opposed to the initial states, $\bigl|\phi_{i}^{+}\bigr\rangle$
and $\bigl|\phi_{i}^{-}\bigr\rangle$, the final state, $\bigl|\phi_{f}\bigr\rangle$,
acts on both forward and backward time-evolution operators.

The partition function assumes a simpler form for initial states defined
by the distribution,
\begin{equation}
\hat{\rho}=\rho^{\hat{b}^{\dagger}\hat{b}}\,,\;\rho=\frac{\bar{n}}{1+\zeta\bar{n}},\label{eq:Initial_Distribution}
\end{equation}
where $\bar{n}$ is the average occupation of the level,
\begin{equation}
\bar{n}=\langle\hat{a}^{\dagger}\hat{a}\rangle_{\rho}=\frac{\mathrm{Tr}\left[\hat{a}^{\dagger}\hat{a}\hat{\rho}\right]}{\mathrm{Tr}\left[\hat{\rho}\right]}\,,\label{eq:average_Number}
\end{equation}
Here, $\bar{n}\geq0$ for bosons and $0\leq\bar{n}\leq1$ for fermions.
An example of a state described by distribution (\ref{eq:Initial_Distribution})
is the thermal state of a non-interacting system, where $\bar{n}=(e^{\frac{\varepsilon-\mu}{T}}-\zeta)^{-1}$,
$\rho=e^{-\frac{\varepsilon-\mu}{T}}$ is the Boltzmann factor, $T$
is the temperature and $\mu$ is the chemical potential. Using distribution
(\ref{eq:Initial_Distribution}) and the identity $\bigl\langle\phi\bigr|\rho^{\hat{b}^{\dagger}\hat{b}}\bigl|\phi'\bigr\rangle=e^{\rho\bar{\phi}\phi'}$
\cite{Kamenev2011}, the integrand in the first line of Eq. (\ref{eq:Partition_Function_1})
becomes $e^{-\left|\phi_{i}^{+}\right|^{2}}\bigl\langle\phi_{i}^{+}\bigr|\hat{\rho}\bigl|\zeta\phi_{i}^{-}\bigr\rangle=e^{-\bar{\phi}_{i}^{+}(\phi_{i}^{+}-\zeta\rho\phi_{i}^{-})}$
and integration over $\bar{\phi}_{i}^{+}$ enforces the constraint
$\phi_{i}^{+}=\zeta\rho\phi_{i}^{-}$. Integrating over $\phi_{i}^{+}$
the partition function (\ref{eq:Partition_Function_1}) becomes,
\begin{equation}
Z=(1-\zeta\rho)^{\zeta}\int\mathrm{d}\bigl[\bar{\phi}_{i},\phi_{i}\bigr]\mathrm{d}\bigl[\bar{\phi}_{f},\phi_{f}\bigr]\,e^{-\bar{\phi}_{i}\phi_{i}}\bigl\langle\phi_{i}\bigr|\hat{\mathcal{U}}\left(t_{i},t_{f}\right)\bigl|\phi_{f}\bigr\rangle\,e^{-\bar{\phi}_{f}\phi_{f}}\bigl\langle\phi_{f}\bigr|\hat{\mathcal{U}}\left(t_{f},t_{i}\right)\bigl|\zeta\rho\phi_{i}\bigr\rangle\,,\label{eq:Partition_Function_2}
\end{equation}
where we remove the label $-$ from the initial state. The pre-factor
of the integral is the normalization of the partition function, $\mathrm{Tr}\left[\hat{\rho}\right]=(1-\zeta\rho)^{\zeta}$.
We express the expectation value of the time-evolution operator in
the functional integral form,
\[
e^{-\bar{\phi}_{f}\phi_{f}}\bigl\langle\phi_{f}\bigr|\hat{\mathcal{U}}\left(t_{f},t_{i}\right)\bigl|\phi_{i}\bigr\rangle=\int_{\phi(t_{i})=\phi_{i}}^{\substack{\phi(t_{f})=\phi_{f}\\
\bar{\phi}(t_{f})=\bar{\phi}_{f}
}
}{\cal D}\bigl[\bar{\phi},\phi\bigr]e^{\mathrm{i}\int_{t_{i}}^{t_{f}}\bar{\phi}(t)\left(\mathrm{i}\partial_{t}-\varepsilon\right)\phi(t)}\,.
\]
Here, the functional integral measure and the action in the exponent
are defined by splitting the time interval $t_{f}-t_{i}$ in $N$
smaller intervals $\delta t=(t_{f}-t_{i})/N$ and taking the symbolical
limit,
\[
\begin{aligned}{\cal D}\bigl[\bar{\phi},\phi\bigr] & =\lim_{N\rightarrow\infty}\prod_{n=1}^{N-1}\mathrm{d}\bigl[\bar{\phi}_{n},\phi_{n}\bigr]\\
\int_{t_{i}}^{t_{f}}\mathrm{d}t\bar{\phi}(t)\left(\mathrm{i}\partial_{t}-\varepsilon\right)\phi(t) & =\lim_{N\rightarrow\infty}\sum_{n=1}^{N}\delta t\left[\mathrm{i}\bar{\phi}_{n}\frac{\phi_{n}-\phi_{n-1}}{\delta t}-\varepsilon\bar{\phi}_{n}\phi_{n-1}\right]
\end{aligned}
\]
where $\phi_{0}=\phi_{i}$ and $\phi_{N}=\phi_{f}$. Integrating over
initial and final fields, we arrive at,
\begin{equation}
Z=(1-\zeta\rho)^{\zeta}\int_{\phi^{+}(t_{i})=\zeta\rho\phi^{-}(t_{i})}^{\phi^{-}(t_{f})=\phi^{+}(t_{f})}\mathcal{D}\bigl[\bar{\phi}^{+},\phi^{+}\bigr]\mathcal{D}\bigl[\bar{\phi}^{-},\phi^{-}\bigr]e^{i\int_{t_{i}}^{t_{f}}\left[\bar{\phi}^{+}(t)\left(\mathrm{i}\partial_{t}-\varepsilon\right)\phi^{+}(t)-\bar{\phi}^{-}(t)\left(\mathrm{i}\partial_{t}-\varepsilon\right)\phi^{-}(t)\right]}\,.\label{eq:Partition_Function_+-}
\end{equation}
where $+$ and $-$ fields are associated to forward and backward
time-evolutions and the minus sign in front of the second term in
the exponent comes from the inversion of the integration limits. The
integration measure has been redefined to contain the additional integration
over the initial and final variables, $\phi_{0}^{-}$, $\bar{\phi}_{0}^{-}$
and $\phi_{N}^{+}$, $\bar{\phi}_{N}^{+}$, which are $\phi^{-}(t_{i})$,
$\bar{\phi}^{-}(t_{i})$ and $\phi^{+}(t_{f})$, $\bar{\phi}^{+}(t_{f})$
in continuum notation. This partition function is the sum over all
the possible configurations of $+$ and $-$ fields, related only
through the boundary conditions in distant past and future, with a
weight that depends on the difference of their actions. The stationary
configurations satisfy $\phi^{+}(t)=\phi^{-}(t)$ and $\bar{\phi}^{+}(t)=\bar{\phi}^{-}(t)$.
These configurations are the most important as, among the individual
configurations, they carry most of the weight of the partition function\footnote{We note that configurations satisfying $\phi^{+}=-\phi^{-}$ and $\bar{\phi}^{+}=-\bar{\phi}^{-}$
are stationary, as well. However, contrary to the previous ones, they
may not be stationary for an interacting system \cite{Kamenev2011}. }. As a consequence, the quantity $\phi^{+}(t)-\phi^{-}(t)$ measures
how far away configurations are from the most important ones.

\subsection*{Bosons}

The goal of the remaining derivation is the evaluation of the Green's
function and the procedure is different for bosons and fermions. We
start from bosons. In analogy to equilibrium quantum field theory
\cite{AltlandSimons2010Book}, we associate stationary configurations,
$\phi^{+}(t)=\phi^{-}(t)$ and $\bar{\phi}^{+}(t)=\bar{\phi}^{-}(t)$,
to classical ones. Motivated by this identification and following
Ref. \cite{Kamenev2011}, we define a classical field, $\phi^{\mathrm{cl}}(t)$,
and a quantum field, $\phi^{\mathrm{q}}(t)$, as
\[
\begin{cases}
\phi^{\mathrm{cl}}(t) & =\frac{1}{\sqrt{2}}\left(\phi^{+}(t)+\phi^{-}(t)\right)\,,\\
\phi^{\mathrm{q}}(t) & =\frac{1}{\sqrt{2}}\left(\phi^{+}(t)-\phi^{-}(t)\right)\,.
\end{cases}
\]
and the same for the complex conjugate. This transformation is known
as Keldysh rotation \cite{Keldysh,Kamenev2011}. In terms of these
new fields, classical configurations have $\phi^{\mathrm{q}}(t)=0$
and, as we will see later, $\phi^{\mathrm{cl}}(t)$ satisfies the
classical equations of motion. Moreover, the quantum field $\phi^{\mathrm{q}}(t)$
measures how far away configurations are from classical ones. In these
new fields, the partition function becomes,
\begin{equation}
Z=\frac{1}{1+\bar{n}}\int_{\phi^{\mathrm{q}}(t_{i})=-\frac{\phi^{\mathrm{cl}}(t_{i})}{1+2\bar{n}}}^{\phi^{\mathrm{q}}(t_{f})=0}\mathcal{D}\bigl[\bar{\phi}^{\mathrm{cl}},\phi^{\mathrm{cl}}\bigr]\mathcal{D}\bigl[\bar{\phi}^{\mathrm{q}},\phi^{\mathrm{q}}\bigr]e^{i\int_{t_{i}}^{t_{f}}\mathrm{d}t\,\bar{\phi}^{\alpha}(t)\left[G_{\bar{n}}^{-1}\right]^{\alpha\beta}(t)\phi^{\beta}(t)}\,,\label{eq:Partition_Function_Bosons}
\end{equation}
where $\alpha,\beta=\mathrm{cl},\mathrm{q}$ and we use the conventional
sum over repeated indices. The inverse Green's function is,
\begin{equation}
\left[G_{\bar{n}}^{-1}\right]^{\alpha\beta}(t)=\left(\begin{array}{cc}
0 & i\partial_{t}-\varepsilon\\
i\partial_{t}-\varepsilon & 0
\end{array}\right)_{\bar{n}}\,.\label{eq:Inverse_Green_Function_Bosons}
\end{equation}
The inverse Green's function depends on the initial average occupation,
$\bar{n}$, because, by construction, it is an operator whose domain
is the set of fields, $\phi^{\alpha}(t)$, that satisfy the boundary
conditions of Eq. (\ref{eq:Partition_Function_Bosons}) and the initial
boundary condition depends on $\bar{n}$. Integration over the quantum
fields, $\phi^{\mathrm{q}}(t)$ and $\bar{\phi}^{\mathrm{q}}(t)$,
enforces the classical equations of motion, $(i\partial_{t}-\varepsilon)\phi^{\mathrm{cl}}(t)=0$
and $(-i\partial_{t}-\varepsilon)\bar{\phi}^{\mathrm{cl}}(t)=0$.
This fact originally motivated the name ``classical field'' \cite{Kamenev2011}.

\begin{figure}
\centering
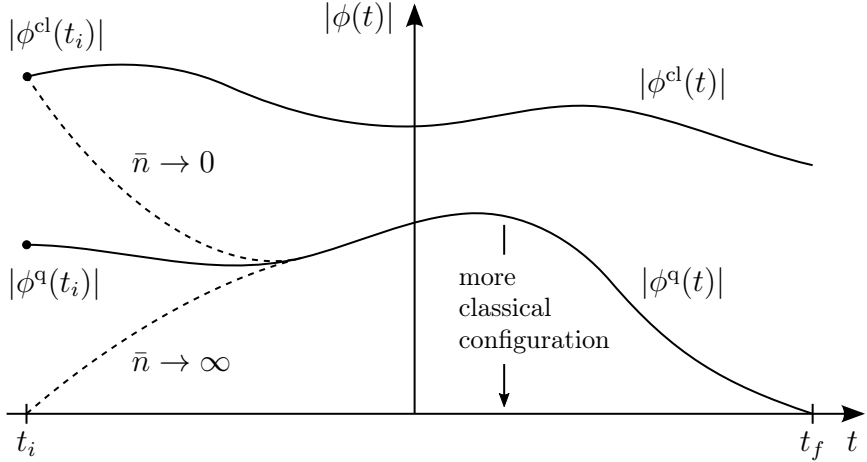

\protect\caption{Example of a configuration of the modulus of quantum and classical
fields, $|\phi^{\mathrm{cl}}(t)|$ and $|\phi^{\mathrm{q}}(t)|$.
The smaller $|\phi^{\mathrm{q}}(t)|$ the closer is the configuration
to a classical one. At the final time, $t=t_{f}$, $|\phi^{\mathrm{q}}(t_{f})|=0$.
Instead, at $t=t_{i}$, the initial value of the quantum field, $|\phi^{\mathrm{q}}(t_{i})|$,
is related to the initial value of the classical field, $|\phi^{\mathrm{cl}}(t_{i})|$,
through Eq. (\ref{eq:Initial_Boundary_Condition}). For a fixed value
of $|\phi^{\mathrm{cl}}(t_{i})|$, $|\phi^{\mathrm{q}}(t_{i})|\rightarrow|\phi^{\mathrm{cl}}(t_{i})|$
for $\bar{n}\rightarrow0$ and $|\phi^{\mathrm{q}}(t_{i})|\rightarrow0$
for $\bar{n}\rightarrow\infty$.\label{fig:Figure}}
\end{figure}

It is worth to pause for a moment and discuss the initial boundary
condition in the partition function (\ref{eq:Partition_Function_Bosons}).
This condition relates the values of initial classical and quantum
variables, $\phi^{\mathrm{cl}}(t_{i})$ and $\phi^{\mathrm{q}}(t_{i})$,
and initial occupation number, $\bar{n}$. Taking the modulus, the
equation reads,
\begin{equation}
|\phi^{\mathrm{q}}(t_{i})|=\frac{\text{1}}{1+2\bar{n}}|\phi^{\mathrm{cl}}(t_{i})|\,.\label{eq:Initial_Boundary_Condition}
\end{equation}
Since for bosons $\bar{n}\geq0$, for a fixed $\phi^{\mathrm{cl}}(t_{i})$
the quantum variable is initially bounded, $0\leq|\phi^{\mathrm{q}}(t_{i})|\leq|\phi^{\mathrm{cl}}(t_{i})|$
(A more restrictive condition applies in the final state, $\phi^{\mathrm{q}}(t_{f})=0$).
When $\bar{n}\ll1$, the level occupation is microscopic and $\phi^{\mathrm{q}}(t_{i})\approx\phi^{\mathrm{cl}}(t_{i})$.
Instead, when $\bar{n}\gg1$, the level occupation becomes macroscopic
and $|\phi^{\mathrm{q}}(t_{i})|\ll|\phi^{\mathrm{cl}}(t_{f})|$. This,
as expected \cite{AltlandSimons2010Book}, means that the larger the
initial occupation of the level, the closer the initial state of the
system is to a classical one. These results are summarized in Fig.
\ref{fig:Figure}.

We proceed and calculate the Green's function, $G^{\alpha\beta}(t,t')=-\mathrm{i}\langle\phi^{\alpha}(t)\bar{\phi}^{\beta}(t')\rangle$,
which satisfies the differential equation,
\[
\left(\begin{array}{cc}
0 & i\partial_{t}-\varepsilon\\
i\partial_{t}-\varepsilon & 0
\end{array}\right)_{\bar{n}}\left(\begin{array}{cc}
G^{\mathrm{cl}\mathrm{cl}}(t,t') & G^{\mathrm{cl}\mathrm{q}}(t,t')\\
G^{\mathrm{q}\mathrm{cl}}(t,t') & G^{\mathrm{q}\mathrm{q}}(t,t')
\end{array}\right)=\left(\begin{array}{cc}
\delta(t-t') & 0\\
0 & \delta(t-t')
\end{array}\right)\,,
\]
where we used the matrix representation (\ref{eq:Inverse_Green_Function_Bosons})
of the inverse Green's function. The general solution of this equation
is,
\begin{equation}
G^{\alpha\beta}(t,t')=-\mathrm{i}e^{-\mathrm{i}\varepsilon(t-t')}\left(\begin{array}{cc}
a & \left(b+1\right)\theta\left(t-t'\right)+b\theta\left(t'-t\right)\\
\left(c+1\right)\theta\left(t-t'\right)+c\theta\left(t'-t\right) & d
\end{array}\right)\,.\label{eq:Green_Function_Bosons_Unknown_Const}
\end{equation}
The constants $a$, $b$, $c$ and $d$ are fixed by the boundary
conditions in (\ref{eq:Partition_Function_Bosons}). For this, we
multiply each boundary condition of the partition function (\ref{eq:Partition_Function_Bosons})
by $\phi^{\alpha}(t')$, take the average and use the definition of
Green's function, $G^{\alpha\beta}(t,t')=-\mathrm{i}\langle\phi^{\alpha}(t)\bar{\phi}^{\beta}(t')\rangle$,
to obtain the boundary conditions of the Green's function,
\begin{equation}
\begin{aligned}G^{\mathrm{q}\mathrm{\alpha}}(t_{f},t') & =0\,,\\
G^{\mathrm{cl}\alpha}(t_{i},t') & =-(1+2\bar{n})G^{\mathrm{q}\alpha}(t_{i},t')\,.
\end{aligned}
\label{eq:Boundary_Conditions_Green_Function_Bosons}
\end{equation}
The first boundary condition fixes the constants of $G^{\mathrm{q}\alpha}(t,t')$
to $c=-1$ for $\alpha=\mathrm{cl}$ and $d=0$ for $\alpha=\mathrm{q}$,
where we use the fact that the first and second theta functions in
$G^{\mathrm{q}\mathrm{cl}}(t_{f},t')$ are respectively equal to 1
and 0 as $t_{i}<t'<t_{f}$. Using $G^{\mathrm{q}\alpha}(t_{i},t')$
in the second boundary condition, the constants of $G^{\mathrm{cl}\alpha}(t,t')$
are fixed to $a=1+2\bar{n}$ for $\alpha=\mathrm{cl}$ and $b=0$
for $\alpha=\mathrm{q}$.\footnote{To fix the constants we did not consider $t'=t_{i},t_{f}$ because
this would lead to equal time Green's functions and the theta functions
in $G^{\mathrm{q}\mathrm{cl}}(t,t')$ and $G^{\mathrm{cl}\mathrm{q}}(t,t')$
have not been regularized at equal times, $t=t'$, yet.} Substituting these values in Eq. (\ref{eq:Green_Function_Bosons_Unknown_Const}),
the Green's function becomes,
\begin{equation}
G^{\alpha\beta}(t,t')=\left(\begin{array}{cc}
G^{\mathrm{K}}(t,t') & G^{\mathrm{R}}(t,t')\\
G^{\mathrm{A}}(t,t') & 0
\end{array}\right)\,,\label{eq:Green_Function_Bosons}
\end{equation}
where,
\begin{equation}
\begin{aligned}G^{\mathrm{R}}(t,t') & =-\mathrm{i}\theta\left(t-t'\right)e^{-\mathrm{i}\varepsilon\left(t-t'\right)}\,,\\
G^{\mathrm{A}}(t,t') & =\mathrm{i}\theta\left(t'-t\right)e^{-\mathrm{i}\varepsilon\left(t-t'\right)}\,,\\
G^{\mathrm{K}}(t,t') & =-\mathrm{i}\left(1+2\bar{n}\right)e^{-\mathrm{i}\varepsilon\left(t-t'\right)}\,.
\end{aligned}
\label{eq:RAK_Green_Functions_Bosons}
\end{equation}
The Green's functions (\ref{eq:RAK_Green_Functions_Bosons}) are retarded,
advanced and Keldysh Green's functions, in agreement with Ref. \cite{Kamenev2011}
for a thermal occupation of the level, $\bar{n}=(e^{-\frac{\varepsilon-\mu}{T}}-1)^{-1}$.
Retarded and advanced Green's functions are still undefined at equal
times because the theta function, $\theta(t)$, needs to be regularized
at $t=0$. First, using the Fourier transform, $G^{R,A}(\omega)=\left(\omega-\omega_{0}+\mathrm{i}0^{\pm}\right)^{-1}$,
we have,
\[
G^{\mathrm{R}}(t,t)-G^{\mathrm{A}}(t,t)=\int\frac{\mathrm{d}\omega}{2\pi}\left[G^{\mathrm{R}}(\omega)-G^{\mathrm{A}}(\omega)\right]=-\mathrm{i}\,.
\]
Second, $\langle\phi^{\mathrm{cl}}(t)\bar{\phi}^{\mathrm{q}}(t)\rangle=-\langle\phi^{\mathrm{q}}(t)\bar{\phi}^{\mathrm{cl}}(t)\rangle^{\dagger}$
which leads to $G^{\mathrm{R}}\left(t,t\right)=\left[G^{\mathrm{A}}\left(t,t\right)\right]^{\dagger}$,
where $\dagger$ denotes complex conjugation and time transposition.
It follows that the regularization is $\theta(t=0)=1/2$.

\subsection*{Fermions}

The derivation is different for fermions, since they do not have a
classical analogue. Instead of classical and quantum, we use the labels
$1$ and $2$ through the transformation \cite{Kamenev2011},
\[
\begin{array}{cc}
\begin{cases}
\phi_{1}(t) & =\frac{1}{\sqrt{2}}\left(\phi^{+}(t)+\phi^{-}(t)\right)\\
\phi_{2}(t) & =\frac{1}{\sqrt{2}}\left(\phi^{+}(t)-\phi^{-}(t)\right)\,,
\end{cases} & \quad\begin{cases}
\bar{\phi}_{1}(t) & =\frac{1}{\sqrt{2}}\left(\bar{\phi}^{+}(t)-\bar{\phi}^{-}(t)\right)\\
\bar{\phi}_{2}(t) & =\frac{1}{\sqrt{2}}\left(\bar{\phi}^{+}(t)+\bar{\phi}^{-}(t)\right)\,.
\end{cases}\end{array}
\]
In these new variables, the partition function reads,
\begin{equation}
Z=\frac{1}{1+\bar{n}}\int_{\phi_{1}(t_{i})=-(1-2\bar{n})\phi_{2}(t_{i})}^{\phi_{2}(t_{f})=0}\mathcal{D}\bigl[\bar{\phi}_{1},\phi_{1}\bigr]\mathcal{D}\bigl[\bar{\phi}_{2},\phi_{2}\bigr]e^{i\int_{t_{i}}^{t_{f}}\mathrm{d}t\,\bar{\phi}_{a}(t)\left[G_{\bar{n}}^{-1}\right]^{ab}(t)\phi_{b}(t)}\,,\label{eq:Partition_Function_Fermions}
\end{equation}
where $a,b=1,2$ and the inverse Green's function is,
\begin{equation}
\left[G_{\bar{n}}^{-1}\right]^{ab}(t)=\left(\begin{array}{cc}
i\partial_{t}-\varepsilon & 0\\
0 & i\partial_{t}-\varepsilon
\end{array}\right)_{\bar{n}}\,.\label{eq:Inverse_Green_Function_Fermions}
\end{equation}
The Green's function, $G^{ab}(t,t')=-\mathrm{i}\langle\phi^{a}(t)\bar{\phi}^{b}(t')\rangle$,
satisfies the differential equation,
\[
\left(\begin{array}{cc}
i\partial_{t}-\varepsilon & 0\\
0 & i\partial_{t}-\varepsilon
\end{array}\right)_{\bar{n}}\left(\begin{array}{cc}
G^{11}(t,t') & G^{12}(t,t')\\
G^{21}(t,t') & G^{22}(t,t')
\end{array}\right)=\left(\begin{array}{cc}
\delta(t-t') & 0\\
0 & \delta(t-t')
\end{array}\right)\,,
\]
for which the general solution is,
\begin{equation}
G^{ab}(t,t')=-\mathrm{i}e^{-\mathrm{i}\varepsilon(t-t')}\left(\begin{array}{cc}
\left(a+1\right)\theta\left(t-t'\right)+a\theta\left(t'-t\right) & b\\
c & \left(d+1\right)\theta\left(t-t'\right)+d\theta\left(t'-t\right)
\end{array}\right)\,,\label{eq:Green_Function_Fermions_Unknown_Const}
\end{equation}
We fix the constants $a$, $b$, $c$ and $d$ using the boundary
conditions of the Green's function,
\begin{equation}
\begin{aligned}G^{2a}(t_{f},t') & =0\,,\\
G^{1a}(t_{i},t') & =-(1-2\bar{n})G^{2a}(t_{i},t')\,.
\end{aligned}
\label{eq:Boundary_Conditions_Green_Function_Fermions}
\end{equation}
obtained from the partition function (\ref{eq:Partition_Function_Fermions})
in the same way we did for bosons. The boundary conditions at $t=t_{f}$
fix $c=0$ and $d=-1$ and the ones at $t=t_{i}$ fix $a=0$ and $b=1-2\bar{n}$.
Substituting these values in Eq. (\ref{eq:Green_Function_Fermions_Unknown_Const}),
the Green's function becomes,
\begin{equation}
G^{ab}(t,t')=\left(\begin{array}{cc}
G^{\mathrm{R}}(t,t') & G^{\mathrm{K}}(t,t')\\
0 & G^{\mathrm{A}}(t,t')
\end{array}\right)\,,\label{eq:Green_Function_Fermions}
\end{equation}
where,
\begin{equation}
\begin{aligned}G^{\mathrm{R}}(t,t') & =-\mathrm{i}\theta\left(t-t'\right)e^{-\mathrm{i}\varepsilon\left(t-t'\right)}\,,\\
G^{\mathrm{A}}(t,t') & =\mathrm{i}\theta\left(t'-t\right)e^{-\mathrm{i}\varepsilon\left(t-t'\right)}\,,\\
G^{\mathrm{K}}(t,t') & =-\mathrm{i}\left(1-2\bar{n}\right)e^{-\mathrm{i}\varepsilon\left(t-t'\right)}\,.
\end{aligned}
\label{eq:RAK_Green_Functions_Fremions}
\end{equation}
As in the bosonic case, we have retarded, advanced and Keldysh Green's
function, in agreement with Ref. \cite{Kamenev2011} for a thermal
occupation of the level, $\bar{n}=(e^{-\frac{\varepsilon-\mu}{T}}+1)^{-1}$.
The regularization of the theta functions are the same as for bosons.
The only difference between Eqs. (\ref{eq:RAK_Green_Functions_Bosons})
and (\ref{eq:RAK_Green_Functions_Fremions}) is the minus sign in
front of $\bar{n}$ in the Keldysh Green's function.

\subsection*{Many levels}

In the previous pages we derived the Green's function for systems
with a single level. In this section, we extend the results to systems
with many levels. In a system with many levels, for every level $i$,
we consider the annihilation and creation operators $\hat{a}_{i}$
and $\hat{a}_{i}^{\dagger}$, satisfying the commutation or anti-commutation
relations $[\hat{a}_{i},\hat{a}_{j}^{\dagger}]_{\pm}=\delta_{ij}$.
The single-level non-interacting Hamiltonian $H=\varepsilon\hat{a}^{\dagger}\hat{a}$,
is replaced by $H=\hat{a}_{i}^{\dagger}\varepsilon_{ij}\hat{a}_{j}$
where, $\varepsilon_{ij}$, is a Hermitian matrix and the average
particle number, $\bar{n}$, is replaced by the one-body density matrix
$\bar{n}_{ij}=\langle\hat{a}_{i}^{\dagger}\hat{a}_{j}\rangle$, a
Hermitian matrix, as well. For clarity, we employ the bold notation
for vectors, $\hat{\boldsymbol{a}}=(\dots,\,\hat{a}_{i-1},\,\hat{a}_{i},\,\hat{a}_{i+1},\,\dots)$,
and matrices, $\boldsymbol{\varepsilon}$ and $\bar{\boldsymbol{n}}$.
To proceed, we diagonalize the density matrix through a unitary transformation,
so that the initial distribution $\hat{\rho}$ is the product of single-level
distributions (\ref{eq:Initial_Distribution}), derive the partition
function and transform back to the original density matrix. Associating
the coherent states vectors $\boldsymbol{\phi}$ and $\bar{\boldsymbol{\phi}}$
to the operators $\hat{\boldsymbol{a}}$ and $\hat{\boldsymbol{a}}^{\dagger}$,
the partition function of bosons and fermions are respectively (\ref{eq:Partition_Function_Bosons})
and (\ref{eq:Partition_Function_Fermions}) with bold notation and
boundary conditions given by, 
\[
\begin{cases}
\boldsymbol{\phi}^{\mathrm{q}}(t_{f})=0\,,\\
\boldsymbol{\phi}^{\mathrm{q}}(t_{i})=-(\boldsymbol{I}+2\bar{\boldsymbol{n}}^{\mathrm{T}})^{-1}\boldsymbol{\phi}^{\mathrm{cl}}(t_{i})\,,
\end{cases}
\]
and
\[
\begin{cases}
\boldsymbol{\phi}_{2}(t_{f})=0\,,\\
\boldsymbol{\phi}{}_{1}(t_{i})=-(\boldsymbol{I}-2\bar{\boldsymbol{n}}^{\mathrm{T}})\boldsymbol{\phi}_{2}(t_{i})\,,
\end{cases}
\]
where $\boldsymbol{I}$ is the identity matrix over the many level
indices and $\mathrm{T}$ denotes transposition. The procedure to
calculate the Green's function is the same as before with the difference
that the columns on which we apply the boundary conditions are over
the many-level indices, as well. The Green's function are identical
in form to (\ref{eq:Green_Function_Bosons}) for bosons and (\ref{eq:Green_Function_Fermions})
for fermions, but now each component is a matrix over the many-level
indices given by, 
\[
\begin{aligned}\boldsymbol{G}^{R}(t,t') & =-\mathrm{i}\theta\left(t-t'\right)e^{-\mathrm{i}\boldsymbol{\varepsilon}\left(t-t'\right)}\,,\\
\boldsymbol{G}^{A}(t,t') & =\mathrm{i}\theta\left(t'-t\right)e^{-\mathrm{i}\boldsymbol{\varepsilon}\left(t-t'\right)}\,,\\
\boldsymbol{G}^{K}(t,t') & =-\mathrm{i}e^{-\mathrm{i}\boldsymbol{\varepsilon}(t-t_{i})}\left(\boldsymbol{I}+2\zeta\bar{\boldsymbol{n}}^{\mathrm{T}}\right)e^{\mathrm{i}\boldsymbol{\varepsilon}(t'-t_{i})}\,.
\end{aligned}
\]
The transposition of the many-body density matrix is due to the fact
that the order of creation and annihilation operator in the definitions
of Green's function and one-body density matrix are exchanged.

\subsection*{Conclusions}

We saw how to derive the non-equilibrium functional integral for bosons
and fermions by accounting for the initial distribution through the
boundary conditions of the functional integral and without resorting
to a discrete representation to find the Green's function. Using this
approach, not only is the derivation shorter and simpler than the
discrete approach but highlights the properties of the classical configurations.
Moreover, it is easy to extend the procedure to many-level systems
with non-diagonal density matrices.

\section{Acknowledgements}

I am grateful to D. M. Gangardt, A. J. Kingl and M. Jones for helpful
discussions. I acknowledge the support of the University of Birmingham.

\bibliographystyle{unsrt}
\bibliography{Bibliography}

\end{document}